\documentclass[letter,twocolumn]{jpsj2}
\usepackage{epsfig}

\title{
Phase Field Model for Dynamics of Sweeping Interface
}

\author{Takuya \textsc{IWASHITA}$^{1}$, Yumino \textsc{HAYASE}$^{2}$,
 and Hiizu \textsc{NAKANISHI}$^{1}$}

\inst{
$^{1}$Department of Physics, Kyushu University 33, Fukuoka 812-8581 \\
$^{2}$Research Institute for Electronic Science, Hokkaido University,
Sappro 060-0812
}

\recdate{\today}

\abst{ Motivated by the drying pattern experiment by Yamazaki and
Mizuguchi[J. Phys. Soc. Jpn. {\bf 69} (2000) 2387], 
we propose the dynamics of sweeping interface, in which
material distributed over a region is swept by a moving interface.  A
model based on a phase field is constructed and results of numerical
simulations are presented for one and two dimensions.  Relevance of the
present model to the drying experiment is discussed.  }

\kword{
interface dynamics, sweeping interface, pattern formation, phase field
model, granular system
}

\makeatletter
\def\lsim{\mathrel{\mathpalette\gl@align<}}
\def\gsim{\mathrel{\mathpalette\gl@align>}}
\def\gl@align#1#2{\lower.6ex\vbox{\baselineskip\z@skip\lineskip\z@
    \ialign{$\m@th#1\hfil##\hfil$\crcr#2\crcr\sim\crcr}}}
\makeatother

\begin {document}
\sloppy
\maketitle

Yamazaki and Mizuguchi\cite{YM00} have shown that fascinating labyrinthine
patterns are formed simply by drying the mixture of water and corn
starch powder that is confined in the two-dimensional space between
two glass plates;  During the drying process, 
the powder is trapped at the interface due to the surface tension and
swept away, but the interface
becomes corrugated and leaves intricate patterns of powder behind.

This process poses a new type of interface dynamics, which we call {\em
the sweeping interface dynamics}; As the interface passes over the space
where some material, e.g. the powder in the above experiment, is
distributed, it sweeps and collects the material along itself.  If the
accumulated material resists to move due to friction or some other
mechanism, the part of the interface where the material density is
larger than at neighboring parts falls behind, then more material will
be swept into the region from the neighbor.  This process causes the
instability of the flat interface.

In this report, in order to study the dynamics of sweeping interface, we
construct a model based on a phase field and examine its behavior by
numerical simulations.


The model consists of two field variables, the phase field $u$ and the
material density $v$, in the two dimensional space.
The interface dynamics is simulated by the time-dependent
Ginzburg-Landau equation for the phase field $u(x,y,t)$,
\begin{equation}
\tau{\partial u\over \partial t}
 = \ell^2\nabla^2 u + u(1-u)\left(u-{1\over 2}-b\right) ,
\label{u-eq}
\end{equation}
with a constant $b$;
$\tau$ and $\ell$ are the time and the length scales, respectively, that
characterize the interface.
Its RHS may be derived from the free energy potential.
We assume $0<b<1/2$, then the system has the stable phase with $u=0$ and
the meta stable phase with $u=1$, which we assign for the dry and the
wet phase, respectively, in the present context.  It is easy to see that
eq.(\ref{u-eq}) has the solution
\begin{equation}
u(x,y,t) = {1\over 2}\left[
1+\tanh\left({1\over\sqrt{8}\, \ell}
\left( x-\sqrt 2 b\,{\ell\over\tau}\,t
\right)\right)\right] ,
\end{equation}
which represents the flat interface between the two phases and it is
advancing to expand the stable dry phase with the speed proportional to $b$.

As for the material density field $v(x,y,t)$, the total amount of
material should be conserved, therefore, the density field follows the
equation of continuity
\begin{equation}
{\partial v \over \partial t} = -\mib{\nabla} \cdot \mib{J}
\label{v-eq}
\end{equation}
with $\mib{J}$ being the flux of material flow.

Sweeping by the interface is represented by the material flux given by
\begin{equation}
\mib{J} = \left(A(v)\mib{\nabla}u\right) v - D(u)\mib{\nabla}v .
\label{flux}
\end{equation}
The first term of RHS represents the flux driven by the interface and
the second term is the diffusion flux.  In the first term, the factor
$(A(v)\mib{\nabla}u)$ can be regarded as the flowing speed of the
material driven by the interface, then the factor $\mib{\nabla} u$
represents the driving intensity and $A(v)$ is the factor proportional
to the mobility of the matter.  The mobility $A(v)$ should be a
decreasing function because, in the region with higher density, the material is
difficult to flow due to the friction effect.
We also introduce the diffusion only within the interface region by assuming
\begin{equation}
D(u) = D_0 u(1-u)
\end{equation}
with a constant $D_0$.
This diffusion flux is originally introduced to eliminate a small scale
instability, which appears in numerical simulations without the
diffusion, but it is plausible that some diffusion process always exists
in a real system.

As a reaction to the driving force on the material, the interface slows
down in the large $v$ region.  This may be represented by making a
constant $b$ in eq.(\ref{u-eq}) a decreasing function of $v$.

The central part of the present model to embody the sweeping dynamics is
in the interaction between the interface in the phase field and the
material density given by the first term of $\mib{J}$ in eq.(\ref{flux})
and $b(v)$ in eq.(\ref{u-eq}), but their detailed
forms are rather arbitrary.  The important points for the sweeping
dynamics of propagating interface are
(i) the interface drives the material flux that is normal to the
interface,
(ii) the friction of material resists the drive
(iii) the reaction of the drive from the material slows down the
interface propagation upon increasing the material density.

We employ simple forms for $b(v)$ and $A(v)$,
\begin{equation}
b(v) = {b_0\over v/v_b +1},
\qquad
A(v) = {A_0\over v/v_A + 1},
\label{b-A}
\end{equation}
with constants $b_0$, $A_0$, $v_b$, and $v_A$.


The initial material distribution is assumed to be uniform with a
constant density $v_0$, namely, $v(x,y,t=0)=v_0$, and the simulations
start with a phase field
\begin{equation}
u(x,y,t=0) = 
{1\over 2}\Bigl[ 1+ \tanh\left({1\over\ell}(x-x_i(y))\right) \Bigr]
\label{u-init}
\end{equation}
with an initial interface position $x_i(y)$.  Numerical integration is
performed with the spatial step $\Delta x=\Delta y=0.5$ and the time
step $\Delta t=0.03$, using the central differences for the spatial
derivatives, and Euler method for the time derivative in the phase field
equation (\ref{u-eq}) and CIP (Cubic Interpolated Profile)
method\cite{CIP1,CIP2} for the equation of continuity (\ref{v-eq}).  The
following simulations are limited to the case of $v_A=v_b$, and we
employ the unit system where $v_b$, $\tau$, and $\ell$ are unities.
Most of simulations are done with the parameters $A_0=8$, $b_0=0.4$, and
$D_0=1$ with $v_0=0.5$, unless stated otherwise.


\begin{figure}[b]
\begin{center}
\epsfig{width=8.7cm,file=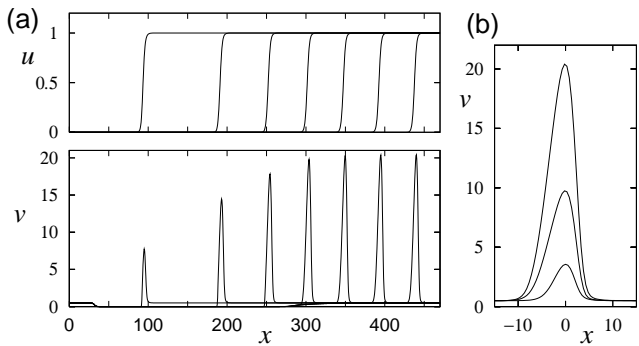}
\end{center}
\caption{ (a) The time series of 1-d system.  The upper and lower graphs
show $u$ and $v$ fields, respectively and the lines with the leftmost
step and the peak are $u$ and $v$, respectively, at $t=300$, and the
following plots are with the time interval 1200 from left to right.  The
parameters $A_0=8$, $b_0=0.4$, $D_0=1$, and $v_0=0.5$.  (b) The steady
profiles of the material density $v$ for ($A_0$, $b_0$)= (8, 0.4),
(4,0.2), and (4,0.4) from top to bottom.  The other parameters are
$D_0=1$ and $v_0=0.5$.  } \label{Fig-1}
\end{figure}

First, we examine the one-dimensional solution, which might be
considered to be the solution uniform in the $y$ direction in two
dimensions; The initial $u$ is the flat interface with constant $x_i$ in
eq.(\ref{u-init}).  Fig.1(a) shows the time development of the system in
one dimension; Initially, the material piles up around the interface and
the peak of $v$ increases as the interface advances, but eventually the
peak in $v$ becomes stationary when the interface passes around
$x\approx 300$ because the material starts overflowing behind the
sweeping interface.  The speed of the interface slows down due to the
friction as the material is accumulated, but the system reaches the
steady state, where the profile of $v$ is stationary in time.  The
stationary profiles are shown for a few sets of $A_0$ and $b_0$ in
Fig.1(b); One can see the peak of $v$ is large for large $A_0$ or for
small $b_0$.  In the steady propagation, the material density behind the
interface is same with the density ahead of it even though the flux is
not zero within the interface; All the material is simply displaced by a
certain distance.

The behavior without the diffusion flux in one dimension is shown in
Fig.2, where the profiles of $u$ and $v$ are plotted in the case of
$D_0=0$. 
Without the diffusion, one can see that $v$ develops a discontinuity
in the rear side of the interface
down to the finest spatial step introduced in numerics.  We have tried a
few functional forms for $A(v)$ and $b(v)$, and parameter values,
but in all the cases, $v$ eventually develops similar discontinuity when
we uses a parameter set that yields substantial accumulation.

This instability is similar to the one in Burgers equation in the
inviscid limit; In the rear side of the interface, the density $v$ is
smaller than that in the preceding part.  In the small $v$ region, the
material flowing speed is large because $A(v)$ is large for small $v$,
therefore, $v$ in the rear side is pushed forward to make a steep cliff.

\begin{figure}[b]
\begin{center}
\epsfig{width=7cm,file=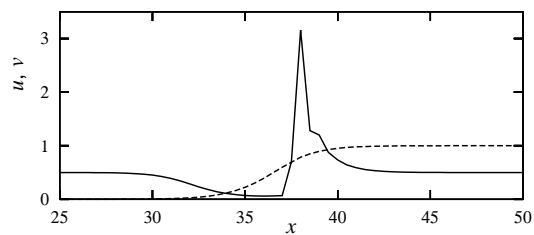}
\end{center}
\caption{
The small scale singularity in $v$ developed in the case of $D_0=0$
with $A_0=8$, $b_0=0.4$, and $v_0=0.5$.
The solid and dashed lines represent $v$ and $u$, respectively.
}
\label{Fig-2}
\end{figure}

Now, we study the two-dimension system.  In two dimensions, the
propagation of a long straight sweeping interface is unstable;  If a
disturbance makes accumulated material density at a certain part along
the interface larger than that at neighbors, then the advancing speed of
the interface becomes slower there, thus the interface becomes
concave.  This makes the density $v$ increase further because the material is
swept into the concave part, from which the material overflows behind
the interface.  This mechanism destabilizes a straight interface to make
a corrugated one, which leaves some pattern in the material density
after it passes.

The two-dimensional simulations are done in the system size of $L_x\times
L_y$.  We employ the periodic boundary condition in the $y$
direction, but the boundaries are open in the $x$ direction.

Fig.3 shows the time development of the rectangular system with
$L_x\times L_y=250\times25$ from the initial phase field with a
sinusoidally perturbed interface with the period $L_y$,
\begin{equation}
x_i(y)\equiv x_0 + R\cos\left({2\pi\over L_y}y \right),
\label{u-init-2}
\end{equation} 
where $x_0$ is the initial position and $R$ is the amplitude of the
perturbation;  We take $x_0=35$ and $R=5$ with the origin at the lower
left corner.  The left and right columns show the $u$ and $v$ fields,
respectively. 
After a while, the interface will develop into a stationary form
with a peak structure.
One can see from the figure that
the curved interface propagates steadily with the stationary form and it
leaves a streak of the material accumulated behind the region where the
interface falls behind.
It should be noted that the interface in $u$ is fairly flat and does not
stay with the line structure of $v$.

\begin{figure}
\begin{center}
\epsfig{width=8.7cm,file=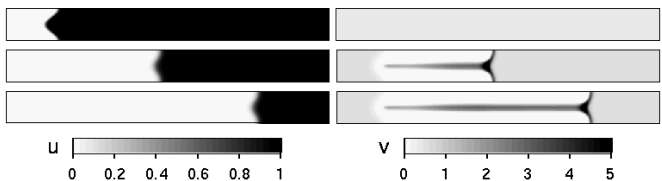}
\end{center}
\caption{
The time development of the 2-d system at $t=0$, 300, 600 from top to
 bottom rows.
The left and right columns show $u$ and $v$, respectively.
the system size is $L_x=250$ and $L_y=25$, and
the parameters are $A_0=8$, $b_0=0.4$, $D_0=1$, and $v_0=0.5$.
}
\label{Fig-3}
\end{figure}

Fig.4 shows snapshots at $t=1800$ for the systems with $L_x=600$
and various width $L_y$;  The initial interface is given by
eqs.(\ref{u-init}) and (\ref{u-init-2}) with a one period for each
$L_y$.  It shows that the curved interface becomes flat for $L_y=15$,
and the single peak interface is unstable against forming another peak
for $L_y\gsim 40$.  The results suggest that there is an optimal spacing
for peaks in the interface.  In the case of the system with $L_y=40$,
the single peak interface is unstable, and another peak tries to emerge,
but cannot make it;  The interface shows a periodic behavior.

\begin{figure}
\begin{center}
\epsfig{width=8.7cm,file=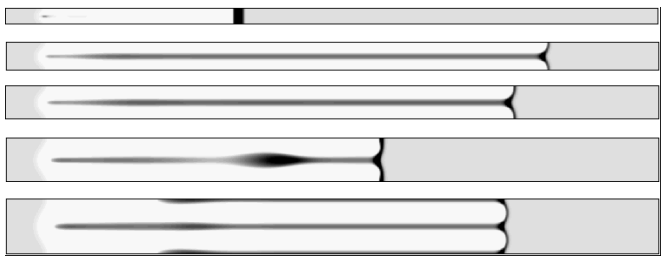}
\end{center}
\caption{ The snapshots of $v$ at $t=1800$ for the system with $L_x=600$
and various width $L_y=$15, 25, 30, 40, and 50 from top to bottom.  The
system length is $L_x=600$ and the other parameters are $A_0=8$,
$b_0=0.4$, $D_0=1$, and $v_0=0.5$.  } \label{Fig-4}
\end{figure}

Fig.5 shows the steady propagation speed of the interface for various
system with $L_y$.  When a curved interface propagates steadily, it is
much faster than the flat interface.  In the curved interface, the
``lighter'' part with less accumulation travels ahead and pull up the
``heavier'' part, from which the material drains behind more efficiently
than in the case of the flat interface.  The speed is slower for a wider
system as long as the interface is single peaked because more material
is accumulated into the concave part of the interface.  For even wider
systems ($L_y\gsim 40$), another peak appears spontaneously, and the
interface speed becomes the same with the system of half width with the
single peak interface.

The steady form of interface changes not only with the system
width $L_y$ but also with the material density $v_0$.  Fig.6 shows
the stability diagram of the curved interface with sweeping pattern;
It can be seen that the curved interface becomes flat for the large
wave number $1/L_y$ and/or the large material density $v_0$.

\begin{figure}
\begin{center}
\epsfig{width=7cm,file=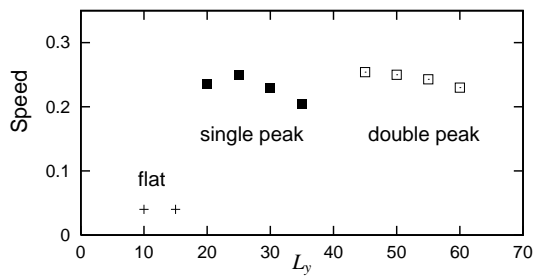}
\end{center}
\caption{
The speed of interface propagation vs. the system width $L_y$. 
The steadily propagating interface are flat (pluses), single peaked (solid
 squares), and double peaked (open squares).
The parameters are
$A_0=8$, $b_0=0.4$, $D_0=1$, and $v_0=0.5$.
}
\label{Fig-5}
\end{figure}

This means that, for larger density of material $v_0$, minimum length
scale of the pattern becomes larger in the present model.  This is
caused by interplay between the friction effect and the
diffusion;  For large $v_0$, the accumulation at the interface
becomes large, which makes the interface motion slow, then the distance
that the material diffuses along the interface becomes large,
and the length scale of the pattern becomes long.

\begin{figure}[b]
\begin{center}
\epsfig{width=7cm,file=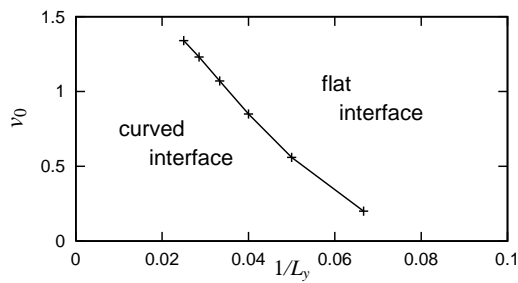}
\end{center}
\caption{
The stability diagram for the curved interface in the $v_0-1/L_y$ plane.
The parameters are
$A_0=8$, $b_0=0.4$, and $D_0=1$.
}
\label{Fig-6}
\end{figure}

Fig.7 is a snapshot of the density field $v$ that emerges from an
initial sinusoidal interface with small irregularity.  The interface is
initially located near the left end of the system, and at the time of
the snapshot, it is located around the edge of the flat $v$ region near
the right end of the system.  One can see that a small perturbation
produces a globally irregular pattern.  If one compare it with the
patterns obtained by Yamazaki and Mizuguchi experiment\cite{YM00}, there
are both similarity and difference; The similarity is that any part of
the region where the material is swept away is connected to the outside
because the pattern is produced by sweeping interface.
The difference is
that the interface remains flat over the scale that spans several
streaks, consequently,
the pattern shows clear direction toward which the interface has
moved, in contrast with the isotropic pattern in the experiment, where
the interface is virtually blocked by clogging powder and corrugated
wildly.  It seems necessary to include more efficient mechanism in the
model to stop the interface propagation than that by the term through
the function $b(v)$ in eqs. (\ref{u-eq}) and (\ref{b-A}).

\begin{figure}
\begin{center}
\epsfig{width=8.7cm,file=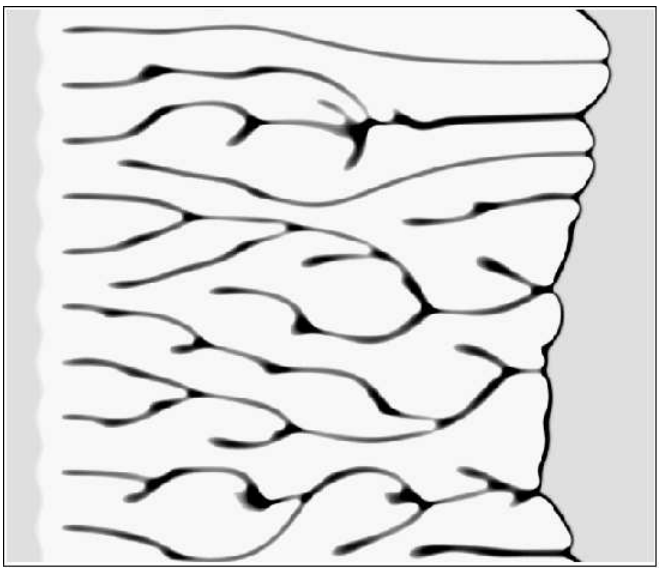}
\end{center}
\caption{
A snapshot of the density $v$ from the initial interface with
a small irregularity.
The system size is $L_x\times L_y=700\times 600$, and
the parameters are
$A_0=8$, $b_0=0.4$, and $D_0=1$ with $v_0=0.5$.
}
\label{Fig-7}
\end{figure}

In the experiment in ref.[1], the gap between the two glass plates is
determined by the largest grain, namely, around 30$\mu{\rm m}$.  The
gap, however, can be controlled by inserting spacers.  It is interesting
to find that the pattern produced by the present model (Fig.7) rather
resembles those found in the experiments with the gap 50$\mu{\rm m}$ or
100$\mu{\rm m}$\cite{Y-private}.  This may be because the friction
effect is smaller in the larger gap plates.

Another difference is that, in the experiment, the scale of pattern does
not depend on the powder density, but in the present simulation, the
length scale becomes large for large $v_0$ due to the diffusion, as we
have seen already.  The diffusion plays a subtle role in the present
model.  It is introduced originally to suppress the small scale
singularity similar to the one in inviscid Burgers equation, but the
diffusion turns out to determine the length scale of the pattern.  In
that respect, we have to examine, for each $v_0$, if there exists the small
$D_0$ parameter region where the small scale singularity is suppressed,
yet the global pattern is not affected by the diffusion.  

In comparison with the phase field model for the crystal
growth\cite{K93}, a major difference is the length scale over which the
coupling field $v$ changes.  In the crystal growth dynamics, the
coupling field is the temperature, and its characteristic length scale
can be much larger than the interface width of the phase field.  Using
this length scale difference, the interface dynamics has been derived
in the narrow interface limit\cite{C89}.  On the other hand, in the
sweeping interface dynamics, a steep change of the coupling field $v$
seems to be intrinsic because the length scale of the variation in $v$
in the normal direction is no longer than the interface width of $u$ even
in the narrow interface limit.

In summary, we propose a dynamics of sweeping interface and
construct a model based on the phase field. Numerical simulations are
performed to find that the flat interface becomes unstable and some
pattern is produced after the interface passes.


The authors thank Dr. Yamazaki for showing unpublished experimental
data, and Dr. Kobayashi for his comments on the phase field model.  This
work is partially supported by a Grant-in-Aid for scientific research
(C) 16540344 from JSPS.


\end {document}